\documentclass[%
 aip,
 jmp,%
 amsmath,amssymb,
preprint,
]{revtex4-1}

\usepackage{graphicx}
\usepackage{dcolumn}
\usepackage{bm}
\usepackage{verbatim}
\usepackage{float}

\begin{document}

\preprint{AIP/123-QED}

\title[B Appelbe et al]{Primary Neutron Spectra in Ion Vlasov-Fokker-Planck Simulations}

\author{B. Appelbe}
\email{bappelbe@ic.ac.uk}
\affiliation{%
 The Centre for Inertial Fusion Studies,  The Blackett Laboratory, Imperial College, London, SW7 2AZ, United Kingdom
}%

\author{W. T. Taitano}
\affiliation{%
Theoretical Division Los Alamos National Laboratory, Los Alamos, New Mexico 87545, USA
 }%

\author{A. J. Crilly}
\affiliation{%
 The Centre for Inertial Fusion Studies,  The Blackett Laboratory, Imperial College, London, SW7 2AZ, United Kingdom
}%

\author{O. M. Mannion}
\affiliation{%
Sandia National Laboratories, Albuquerque, New Mexico 87185, USA
 }%

\author{C. J. Forrest}
\affiliation{%
Laboratory for Laser Energetics, University of Rochester, Rochester, New York 14623, USA
 }%

\author{J. P. Chittenden}
\affiliation{%
 The Centre for Inertial Fusion Studies,  The Blackett Laboratory, Imperial College, London, SW7 2AZ, United Kingdom
}%

\date{\today}

\begin{abstract}
The energy spectra of unscattered neutrons produced by deuterium-deuterium and deuterium-tritium fusion reactions are an important diagnostic in High Energy Density Physics experiments as the spectra are sensitive to the velocities of reacting ions. Methods exist for calculating these spectra in radiation-hydrodynamic\cite{Munro_NF2016} ("hydro") and Particle-in-Cell\cite{Higginson_JCP2019} ("PiC") simulations. The spectra are particularly sensitive to the high energy tail of ion velocity distribution functions since reaction cross sections increase rapidly with the kinetic energy of a reacting ion pair at the energies achieved in laboratory experiments. This means both the hydro and PiC method may not be suitable in certain plasma regimes. The hydro method assumes that the ion velocity distribution is locally Maxwellian, while the PiC method is subject to statistical noise that makes it challenging to accurately simulate finer details of the spectra. In this work, we present a complementary approach: a method for calculating the neutron spectra in ion Vlasov-Fokker-Planck simulations in which the velocity distribution function is fully-resolved. The method is implemented in the spherically-symmetric code iFP\cite{Taitano_POP2018} which is used to simulate laser-driven Inertial Confinement Fusion experiments. The method is computationally intensive as it requires a five-dimensional numerical integral, but no approximations of the distribution functions or differential cross sections are required. Results show that deviations of the ion distribution functions from Maxwellian can have a noticeable effect on neutron spectra in shock-driven ICF implosions. The method should facilitate more accurate benchmarking of simulations and experiments.
\end{abstract}

\maketitle

\section{Introduction}\label{sec:1}
Experiments in High Energy Density Physics (HEDP) involve the creation of plasmas at extreme conditions which change rapidly in both time and space. The scale and complexity of these plasmas means that computer simulations are required for understanding the plasma properties and for designing experiments. A key component of computer simulations is the production of synthetic diagnostics, which are simulated values for the observable quantities that can be measured in experiments. Synthetic diagnostics are important for two reasons: (i) they allow an accurate comparison of simulations and experiments (benchmarking) and (ii) they can identify novel ways to interpret diagnostic data.\cite{Crilly_POP2018,Crilly_POP2020,Crilly_POP2021}

An important diagnostic in HEDP is neutron spectroscopy.\cite{Frenje_PPCF2020,Frenje_NF2013,GatuJohnson_RSI2012} This involves measuring the neutron energy spectra emitted from a plasma. The vast majority of these neutrons are produced by fusion reactions between the deuterium (D) and tritium (T) ions. If these neutrons do not scatter from plasma ions or surrounding hardware before reaching a detector then the neutron energy will be determined by the $Q$-value of the reaction (which is known to high accuracy) and the velocities of the reacting ions. We refer to the energy spectra of unscattered fusion neutrons as the primary neutron spectra, which are the subject of the present work.

The sensitivity of primary neutron spectra to the velocities of reacting ions means that the spectra are a widely used diagnostic of ion kinematics in a plasma. However, the spectra are only representative of those ions which undergo a nuclear reaction. The reaction cross sections for the fusion reactions vary strongly with the kinetic energy of a reacting ion pair (referred to as the reaction energy). In particular, the deuterium-deuterium (DD) and deuterium-tritium (DT) cross sections increase rapidly with increasing reaction energy at the conditions achieved in laboratory experiments. This means spectra shapes are not a uniform diagnostic of all ions but are more sensitive to ions at higher energies in the distribution function.

The goal of primary neutron spectroscopy is usually to infer conditions about the bulk ion population, not just the conditions in the high energy tail. This is exemplified by experiments in Inertial Confinement Fusion (ICF) in which primary neutron spectroscopy is used to infer the ion temperature and bulk fluid velocity of an ICF hotspot.\cite{Mannion_POP2021,GatuJohnson_PRE2016,Murphy_2014} However, such an inference requires either knowledge or an assumption of how the bulk ion distribution is related to the high energy tail component of the distribution in which the reactions take place.

For ICF experiments, the standard approach is to assume the plasma is in a local equilibrium, meaning that it has a Maxwellian distribution function (appropriately shifted by the bulk fluid velocity) at any given point in time and space. The relationship between ion temperature, bulk fluid velocity and spectral shape is well known for a single Maxwellian.\cite{Brysk_1973,Ballabio_1998,Appelbe_2011,Appelbe_HEDP2014} By assuming that a plasma varying in time and space can be modelled as a set of Maxwellians, it is possible to calculate the neutron spectrum emitted by an inhomogeneous plasma. This approach has resulted in a prescription in which the moments of primary neutron spectra can be related to the ion temperatures and bulk fluid velocities found in ICF hotspots.\cite{Munro_NF2016} This prescription is particularly suitable to implementation in radiation-hydrodynamic ("hydro") simulation codes.\cite{Munro_POP2017} The validity of this model is based on the assumption that the ion mean free path $\lambda_{ii}$ is significantly less than the temperature and density scale lengths $l_{T}$ and $l_{n}$. This ensures the local distribution function is close to Maxwellian. However, recent burning plasma experiments on the NIF have found evidence that the distribution functions are non-Maxwellian.\cite{Hartouni_2023}

Computational models for primary neutron spectra for plasmas that are not in a local equilibrium also exist. These include hybrid models,\cite{Appelbe_POP2012,Appelbe_POP2015} in which a population of ions with an arbitrary distribution reacts with a Maxwellian distribution, and Monte Carlo\cite{Knapp_POP2013} and Particle-in-Cell ("PiC") models,\cite{Higginson_JCP2019} in which the distributions of all reactants can be arbitrary. These models are particularly suitable for reactions which occur in plasmas that are far from equilibrium, such as in interpenetrating flows\cite{Higginson_POP2019} and some configurations of Z-pinches\cite{Klir_NJP2020} in which the beam-target mechanism of neutron production dominates. In such situations the neutron spectra can display large energy shifts or line-of-sight variations, allowing the estimation of quantities such as the mean energy of reacting particles or the direction of acceleration of particles. These spectral features can be well characterised by PiC simulations. However, the inherent statistical noise of PiC methods can make it challenging to accurately simulate subtler spectral features, such as the spectral moments that arise in ICF experiments.

In this work we develop a third computational model for primary neutron spectra, namely, the simulation of spectra in ion Vlasov-Fokker-Planck (VFP) simulations. Ion VFP simulations treat the ion distribution function as a continuum in phase space that evolves in time. This means we can use VFP simulations to compute spectra from distribution functions of arbitrary form (we are not limited to the Maxwellian approximation) and without the statistical noise of PiC methods. No approximations of the distribution function calculated by the VFP simulation or of the reaction differential cross section are required to compute the spectra. Therefore, the spectra are an accurate representation of the simulated distribution functions. This has allowed us to investigate how departures of the distribution function from Maxwellian result in observable spectral features.\cite{Mannion_2021}

We note that previous work has been reported\cite{Inglebert_EPL2014} in which primary neutron spectra were computed in ion VFP simulations. However, in that case the distribution functions were firstly approximated as Maxwellian prior to computing spectra. Therefore, the sensitivity of spectra to non-Maxwellian distributions could not be evaluated. As our results demonstrate, non-Maxwellian distributions can have a significant effect on observable spectral features in certain ICF experimental regimes.

The contents of the paper are as follows: In section \ref{sec:3} we give an overview of the iFP code, into which our computational model for spectra has been implemented. In section \ref{sec:4} we give a detailed description of that model and discuss results of the model in section \ref{sec:5}. Finally, section \ref{sec:6} contains conclusions.

\section{The iFP code}\label{sec:3}
Our computational model for primary neutron spectra is implemented in the iFP code (see Taitano et al\cite{Taitano_POP2018,Anderson_JCP2020,Taitano_CPC2021} and references therein). This is an ion VFP code that has been developed for the simulation of imploding spherical capsules used in ICF experiments. It has been particularly useful for understanding the role of strong shocks on ion kinetics, resulting in effects such as species stratification, which occur in many of these experiments.\cite{Taitano_POP2018,Keenan_POP2020,Keenan_POP2018}

The iFP code assumes a spherically symmetric configuration space, that is one spatial dimension. The spatial coordinate is denoted by $r$. Spherical symmetry of configuration space allows us to assume azimuthal symmetry in velocity space. Velocity space co-ordinates are denoted by $v_{\parallel}$, velocity parallel to the radial vector with values $v_{\parallel}\in\left(-\infty,\infty\right)$, and $v_{\perp}$, velocity orthogonal to radial vector with values $v_{\perp}\in\left[0\right.,\left.\infty\right)$. Ion velocity distribution functions can then be denoted by $f_{i}\left(r,\vec{v},t\right)$. The iFP code is capable of multiple ion species, each with a separate distribution function. The code is adaptive in phase space, where the configuration space is adapted based on a nonlinearly stabilized moving mesh partial differential equation (MMPDE) to track macroscopic features such as shocks and material interfaces, while the velocity space expands/contracts and shifts with local and instantaneous values of thermal speed and bulk velocities of individual ion species.\cite{Taitano_CPC2021}

The electrons are treated as a fluid and the plasma is assumed to be quasineutral and ambipolar, allowing the electron density and drift velocity to be calculated. The electric field has both resistive and thermoelectric contributions. A Fokker-Planck collision operator is used to model collisions between ion species and also between each ion species and the fluid electrons. The coupled VFP ion and fluid electron equations are solved fully nonlinearly and implicitly, based on an asymptotic preserving, high-order low-order (HOLO) solver.\cite{Chacon_JCP2017}

\section{A model for primary spectra in iFP}\label{sec:4}
The procedure for calculating primary neutron spectra from an ion VFP code such as iFP is intuitive: at each point in configuration space, we sum up the spectra produced by every possible pair of reactant velocities. For each pair of reactant velocities we need to calculate the energy of neutrons produced in reactions (a function of $Q$ value and reactant velocity vectors) and the number of neutrons produced (a function of particle number density and reaction cross section). In general, this requires a nine dimensional numerical integral, three spatial and six velocity. This can be extremely computationally demanding, particularly since the integration would need to be computed at many times during the simulation.

The spherically-symmetric nature of the iFP code means that we can reduce this to a five dimensional numerical integral, one spatial and five velocity. However, correctly accounting for the spherically-symmetric nature of this problem also requires a correct treatment of the angular emission of neutrons. At a given point in configuration space, the ion distribution functions can be anisotropic in velocity space. Reactions in these distributions would result in anisotropic neutron energy spectra being emitted, and we note that other authors have developed semi-analytic models for such scenarios\cite{Goncharov_2015}. However, the spherical-symmetry of configuration space means that when we spatially-integrate the emitted neutron spectrum over $4\pi$ of configuration space we will obtain an isotropic neutron spectrum. This isotropic spectrum is the same as the spectrum of neutron energies emitted into $4\pi$ by the anisotropic ion distribution function. We exploit this symmetry in our algorithm for neutron spectra by calculating spectra kernels. These kernels are defined to be the energy spectra of neutrons emitted into $4\pi$ by a pair of reacting particles and will depend on both the magnitudes of the velocities of reacting particles and the angle between them. Subsection \ref{sec:4.1} describes the calculation of spectra kernels. Once the kernels are obtained, we can then carry out the integration over phase space, as described in subsection \ref{sec:4.2}.

\subsection{The spectrum kernel for a pair of reactants}\label{sec:4.1}
The spectrum kernel for a reacting pair can be obtained from the kinematic relations for two-body reactions. We denote the reaction by $1+2\rightarrow 3+4$ . We seek the spectrum of energies of particle $3$ (the neutron) for given velocities of particles $1$ and $2$ (the reacting ions). These velocity vectors are denoted $\vec{v}_{1}$ and $\vec{v}_{2}$. The reactions that are of particular interest are the DD and DT neutron producing reactions
\begin{eqnarray}
D+D&&\rightarrow He^{3}+n,\qquad Q\approx 3.27\,MeV,\\
D+T&&\rightarrow He^{4}+n,\qquad Q\approx 17.59\,MeV,
\end{eqnarray}
where $Q$ is the reaction $Q$-value. A trivial change of particle masses, $Q$-values and differential cross sections will allow the method to be applied to other particles and two-body reactions.

We assume non-relativistic kinematics throughout (this is discussed further in section \ref{sec:6}) and so the energy of particle $3$ in the CM frame is defined to be
\begin{eqnarray}\label{e:4.1.1}
E_{3}^{*} &=& \frac{m_{4}}{m_{3}+m_{4}}\left(Q+K\right),\label{e:4.1.1a}\\
K &=& \frac{1}{2}m_{12} v_{r}^{2},\label{e:4.1.1b}
\end{eqnarray}
where $m_{i}$ for $i = 1,\ldots,4$ denotes particle mass, $K$ is reaction energy, $v_{r}=\left|\vec{v}_{1}-\vec{v}_{2}\right|$ is the relative speed between a pair of reactants and $m_{12} = m_{1}m_{2}/\left(m_{1}+m_{2}\right)$ is the reduced mass of the reactants.

We begin the derivation of the spectrum kernel by defining the double-differential cross section for two-body fusion reactions as
\begin{equation}\label{e:4.1.2}
\frac{d\sigma}{d\Omega_{cm}}\delta\left(E_{3}^{'}-E_{3}^{*}\right)d\Omega_{cm}dE_{3}^{'},
\end{equation}
where $d\Omega_{cm} = d\mu_{s}d\phi_{s}$ is the solid angle into which the neutron is emitted in the CM frame, $\frac{d\sigma}{d\Omega_{cm}}$ is the differential cross section in the CM frame, and $E_{3}^{'}$ is a variable representing neutron energy in the CM frame. The differential cross section is a function of $K$ and $\mu_{s}=\cos\theta_{s}$, where $\theta_{s}$ is referred to as the scattering angle. The differential cross section is usually obtained from experimental data. In this work we will use standard forms for differential cross sections,\cite{Drosg_1987} in which the dependence on $\mu_{s}$ is in the form of Legendre polynomials.

Throughout, we denote quantities in the CM frame with a prime $'$, e.g. $\vec{v}_{1}^{'}$. Quantities in the lab frame are unprimed, e.g. $\vec{v}_{1}$. We are seeking the spectrum of neutron energies in the lab frame. Our procedure for calculating the spectrum kernel is to express \eqref{e:4.1.2} as a function of the lab frame neutron energy variable $E_{3}$ and then integrate over all other variables in the expression.

\begin{figure}
\begin{center}%
\includegraphics*[width=0.5\columnwidth]{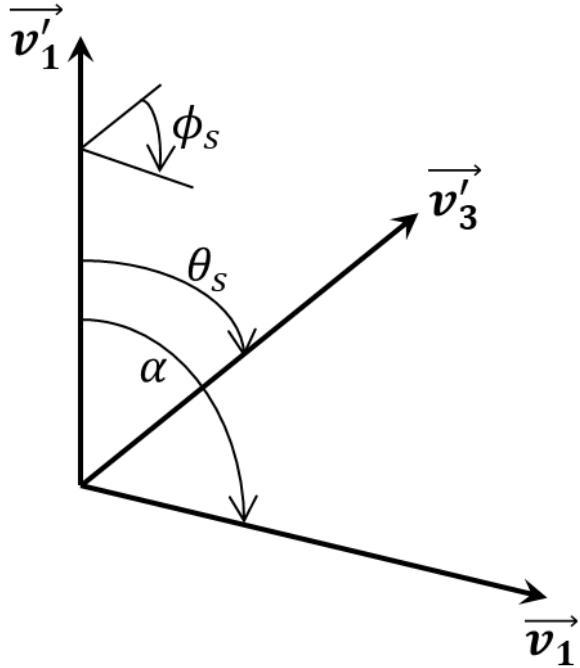}%
\vspace{-1em}
\end{center}
\caption[]{Definition of the angles between vectors $\vec{v}_{1}$, $\vec{v}_{1}^{'}$ and $\vec{v}_{3}^{'}$.} \label{f:1}
\end{figure}

We transform the variables $\left(E_{3}^{'},\phi_{s}\right)\rightarrow\left(E_{3},x\right)$ in \eqref{e:4.1.2} using the relations
\begin{eqnarray}
E_{3} &=& \frac{1}{2}m_{3}\left(\vec{v}_{cm}+\vec{v}_{3}^{'}\right)^{2} ,\label{e:4.1.3a}\\
x &=& E_{3}^{'}-E_{3}^{*},\label{e:4.1.3b}\\
\mu' &=& \frac{1}{v_{cm}}\left(v_{1}\cos\alpha-v_{1}^{'}\right)\mu_{s}+\frac{v_{1}}{v_{cm}}\sqrt{1-\mu_{s}^{2}}\sin\alpha\cos\phi_{s},\label{e:4.1.3c}
\end{eqnarray}
where $\mu'$ represents the cosine of the angle between $\vec{v}_{3}^{'}$ and $\vec{v}_{cm}$, $\alpha$ is the angle between vectors $\vec{v}_{1}^{'}$ and $\vec{v}_{1}$, as shown in fig. \ref{f:1}, and we have used the relation $\vec{v}_{cm} = \vec{v}_{1}-\vec{v}_{1}^{'}$ to obtain \eqref{e:4.1.3c}.

The transformed double-differential cross section can then be expressed as
\begin{equation}\label{e:4.1.4}
\frac{d\sigma}{d\Omega_{cm}}\delta\left(x\right)\frac{\partial \phi_{s}}{\partial x} \frac{\partial E_{3}^{'}}{\partial E_{3}}d\mu_{s}dxdE_{3}\nonumber
\end{equation}
which, after trivial integration over $x$, becomes
\begin{equation}\label{e:4.1.5}
\frac{1}{\sqrt{2m_{3}E_{3}^{*}}}\frac{2}{B\sin\phi_{s}}\frac{d\sigma}{d\Omega_{cm}}d\mu_{s}dE_{3},
\end{equation}
where
\begin{eqnarray}
\sin\phi_{s} &=& \frac{1}{B}\sqrt{B^{2}-\left(C-A\right)^{2}},\label{e:4.1.6a}\\
A &=& \left(v_{1}\cos\alpha-v_{1}^{'}\right)\mu_{s},\label{e:4.1.6b}\\
B &=& v_{1}\sqrt{1-\mu_{s}^{2}}\sin\alpha,\label{e:4.1.6c}\\
C &=&\frac{1}{\sqrt{2m_{3}E_{3}^{*}}}\left(E_{3}-E_{3}^{*}-\frac{1}{2}m_{3}v_{cm}^{2}\right).\label{e:4.1.6d}
\end{eqnarray}
We have included the factor of $2$ in \eqref{e:4.1.5} to account for the fact that $\phi_{s}\rightarrow x$ is multi-valued in the range $\phi_{s}\in\left[0,2\pi\right]$.

The utility of this transformation is that our double-differential cross section is now a function of $E_{3}$, the variable whose spectrum we seek, and $\theta_{s}$, the variable determining the direction of neutron emission.

Now, we want to find the values of $\mu_{s}$ such that the expression for $\sin\phi_{s}$ is real, that is $B^{2}-\left(C-A\right)^{2}\geq 0$. Imaginary values of $\sin\phi_{s}$ correspond to pairs of values $\left(E_{3},\theta_{s}\right)$ that are kinematically impossible to obtain for specified values of $v_{cm}$, $v_{r}$ and $\cos\alpha$. The minimum and maximum permissible values of $\mu_{s}$ are
\begin{equation}
\max\left(-1,r_{-}\right) \leq\mu_{s}\leq \min\left(1,r_{+}\right),\label{e:4.1.7}
\end{equation}
where
\begin{eqnarray}
r_{\pm} &=& \frac{1}{v_{cm}^{2}}\left(v_{1}\cos\alpha-v_{1}^{'}\right)C\pm\frac{v_{1}\sin\alpha}{v_{cm}^{2}\sqrt{2m_{3}E_{3}^{*}}}\sqrt{4E_{3}E_{3}^{*}-\left(E_{3}^{*}+E_{3}-\frac{1}{2}m_{3}v_{cm}^{2}\right)^{2}}.\label{e:4.1.8}
\end{eqnarray}
With these limiting values we can integrate \eqref{e:4.1.5} over $\mu_{s}$ to obtain the spectrum kernel
\begin{eqnarray}
&&\mathcal{K}_{12}\left(E_{3};v_{cm},v_{r},\cos\alpha\right) = \nonumber\\ &&\frac{2}{\sqrt{2m_{3}E_{3}^{*}}}\left[\int_{\mu_{L}}^{\mu_{U}}\frac{1}{\sqrt{B^{2}-\left(C-A\right)^{2}}}\frac{d\sigma}{d\Omega_{cm}}d\mu_{s}\right]dE_{3},\label{e:4.1.9}
\end{eqnarray}
where the limits of integration are $\mu_{L} = \max\left(-1,r_{-}\right) $ and $\mu_{U} = \min\left(1,r_{+}\right)$.

The spectrum kernel $\mathcal{K}_{12}$ represents the spectrum of neutron energies emitted into $4\pi$ for reactions between a pair of particles with velocities $\vec{v}_{1}$ and $\vec{v}_{2}$. The integral over $\mu_{s}$ in \eqref{e:4.1.9} has an analytic solution if the dependence of differential cross section on $\mu_{s}$ can be expressed in the form Legendre polynomials. This solution is given in \ref{app:1}. It means that our spectrum kernel can be expressed as an analytic function of $v_{cm}$, $v_{r}$ and $\cos\alpha$. These quantities are dependent on $v_{1}$ and $v_{2}$ as follows
\begin{eqnarray}
v_{cm} &=& m_{12}\sqrt{\frac{v_{1}^{2}}{m_{2}^{2}}+\frac{v_{2}^{2}}{m_{1}^{2}}+2\frac{\vec{v}_{1}\cdot\vec{v}_{2}}{m_{1}m_{2}}},\label{e:4.1.10a}\\
v_{r} &=& \sqrt{v_{1}^{2}+v_{2}^{2}-2\vec{v}_{1}\cdot\vec{v}_{2}},\label{e:4.1.10b}\\
\cos\alpha &=& \frac{1}{v_{1}v_{r}}\left(v_{1}^{2}-\vec{v}_{1}\cdot\vec{v}_{2}\right).\label{e:4.1.10c}
\end{eqnarray}

Finally, we note that simpler expressions for the spectrum kernel exist when $\sin\alpha = 0$ and $v_{cm} = 0$. These limiting cases are discussed in \ref{app:2}.

\subsection{Integration over phase space}\label{sec:4.2}
Given $\mathcal{K}_{12}$ for an arbitrary pair of reactants, we can now integrate this expression over all of phase space in order to obtain the neutron spectrum emitted from the entire plasma. We begin with the integration over the velocity space. For arbitrary velocity distributions of reactants, $f_{1}\left(r,\vec{v}_{1},t\right)$ and $f_{2}\left(r,\vec{v}_{2},t\right)$, this integral is six-dimensional
\begin{equation}\label{e:4.2.1}
S_{12}\left(E_{3}\right) = \frac{1}{1+\delta_{12}}\frac{1}{4\pi}\int v_{r}\mathcal{K}_{12}f_{1}f_{2}d^{3}\vec{v}_{1}d^{3}\vec{v}_{2},
\end{equation}
where $S_{12}\left(E_{3}\right)$ is the number of neutrons per unit energy per steradian.

In spherically-symmetric geometry the reactant distributions are a function of two co-ordinates, $v_{\parallel}$ and $v_{\perp}$. However, $\mathcal{K}_{12}$ also depends on the angle between $\vec{v}_{1}$ and $\vec{v}_{2}$, which we define as
\begin{eqnarray}
\frac{\vec{v}_{1}\cdot\vec{v}_{2}}{v_{1}v_{2}} &=& \cos\theta_{1}\cos\theta_{2}+\sin\theta_{1}\sin\theta_{2}\cos\phi_{2}\quad,\label{e:4.2.2a}\\
\cos\theta_{1,2} &=& \frac{v_{\parallel 1,2}}{v_{1,2}}\quad,\label{e:4.2.2b} \\
v_{1,2} &=& \sqrt{v_{\parallel 1,2}^{2}+v_{\perp 1,2}^{2}}\quad,\label{e:4.2.2c}
\end{eqnarray}
where $\theta_{1,2}$ are the polar angles between the velocity vectors and the radial vector and $\phi_{2}$ is an azimuthal angle.

Equation \eqref{e:4.2.1} can now be expressed as
\begin{equation}\label{e:4.2.3}
S_{12}\left(E_{3}\right) = \frac{1}{1+\delta_{12}}\frac{1}{2}\int_{0}^{\infty}\int_{0}^{\infty}\int_{-\infty}^{\infty}\int_{-\infty}^{\infty}\left[\int_{0}^{2\pi}v_{r}\mathcal{K}_{12}d\phi_{2}\right]v_{\perp 1}v_{\perp 2}f_{1}f_{2}dv_{\parallel 1}dv_{\parallel 2}dv_{\perp 1}dv_{\perp 2},
\end{equation}
which is a five-dimensional numerical integral. Here we have also carried out a trivial integral over the azimuthal angle $\phi_{1}$ of vector $\vec{v}_{1}$ to gain a factor of $2\pi$ compared with \eqref{e:4.2.1}.

Finally, integration of $S_{12}\left(E_{3}\right)$ over the spatial radial co-ordinate gives the energy spectrum of neutrons emitted from reactions between $f_{1}$ and $f_{2}$
\begin{equation}\label{e:4.2.4}
\langle S_{12}\rangle = \int_{0}^{R} r^{2} S_{12}\left(E_{3}\right)dr,
\end{equation}
in units of $J^{-1}\,s^{-1}\,sr^{-1}$.

\section{Results}\label{sec:5}

The primary spectra model was tested by initializing Maxwellian distributions of deuterium and tritium on the velocity space grid of iFP. The spectra of DD and DT neutrons were then compared with semi-analytic expressions\cite{Appelbe_PPCF2011} for the spectra and were found to be in close agreement with a relative error of less than $0.01$ across $4$ orders of magnitude of intensity. Similar agreement was found when a Maxwellian distribution with an additional drift velocity was initialized in iFP. We note that achieving such accuracy in a PiC model would be challenging due to statistical noise. This demonstrates the suitability of ion VFP simulations for studying subtle changes in neutron spectra.

The velocity space grid resolution used in these tests was the resolution which was previously tuned to ensure accuracy of simulations of capsule implosions.\cite{Taitano_POP2018} A coarser grid can also be used to calculate neutron spectra which remain accurate over $1-2$ orders of magnitude of spectral intensity. This can significantly reduce the runtime of the neutron spectra calculation.

\begin{figure}
\begin{center}%
\includegraphics*[width=0.95\columnwidth]{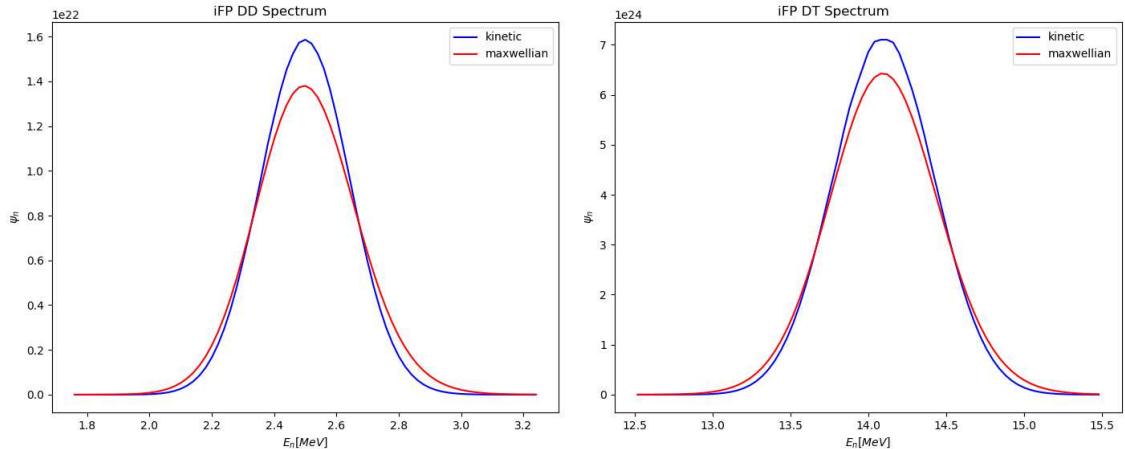}%
\vspace{-1em}
\end{center}
\caption[]{DD (left) and DT (right) neutron spectra at bang time produced from an iFP simulation of a shock driven implosion. The "Kinetic" spectra are those calculated using our model while the "Maxwellian" spectra are calculated by assuming that the ions have a Maxwellian distribution function.} \label{f:2}
\end{figure}

The iFP code with primary spectra model has been used to carry out a series of simulations of spherical shock drive implosions. A detailed analysis of these simulations and the experiments that they accompanied is reported by Mannion et al.\cite{Mannion_2021} Here we select a sample result which illustrates the importance of our new model. Figure \ref{f:2} shows the instantaneous DD and DT neutron spectra produced at bang time in a simulation of a $420\, \mu m$ radius target with a $3\, \mu m$ SiO2 ablator, filled with $5\, atm$ of DT fuel and irradiated by a $0.6\,ns$ square laser pulse. The burn width in the simulation is approximately $150\, ps$. In both cases, two spectra are shown, labelled "kinetic" and "maxwellian". The kinetic spectrum is the result produced from our primary spectra model while the maxwellian spectrum is the result obtained by assuming that the velocity space distribution is Maxwellian at all points in configuration space. At each point in configuration space, the drift velocity and mean kinetic energy of the ions is calculated and this is used to define the temperature and drift velocity of a Maxwellian distribution.

As is clear from fig. \ref{f:2} there are significant differences between the kinetic and maxwellian spectra. These differences can be quantified by calculating an apparent ion temperature from the variance of the spectra.\cite{Ballabio_1998} For DD the kinetic and maxwellian spectra gives values of $16.01\, keV$ and $21.23\, keV$, respectively, while for DT these values are $18.37\, keV$ and $22.49\, keV$. Therefore, a Maxwellian approximation of the distribution functions makes the plasma appear significantly "hotter" by erroneously broadening the neutron spectra.

These differences shows that the neutron spectra shapes are sensitive to details of the ion distribution functions. It is not sufficient to use only the bulk values of these distributions such as drift velocity and mean ion kinetic energy in order to determine spectral shapes. This indicates that neutron spectra can be used as a diagnostic of ion distribution functions, an idea which is validated using experiment and simulations by Mannion et al\cite{Mannion_2021} and explored theoretically in Crilly et al.\cite{Crilly_2021}

\section{Conclusions}\label{sec:6}
We have developed a method for computing primary neutron spectra in spherically-symmetric ion VFP simulations which has been implemented in the iFP code. The method does not require any approximation to the arbitrary distribution functions that can be generated in iFP and so provides a reliable tool for investigating the sensitivity of the experimentally observable neutron spectra to the ion distribution functions. Experiments at the OMEGA facility\cite{Mannion_2021} have demonstrated that this sensitivity provides direct evidence of non-Maxwellian ion distribution functions in shock driven exploding pusher targets and the model developed here means that iFP can more accurately reproduce the experimentally observed spectra than simulations which assume a Maxwellian distribution.

A number of potential minor improvements to the model developed here have been identified. Relativistic kinematics can slightly change the shape of neutron spectra compared with the nonrelativistic model used here.\cite{Appelbe_HEDP2014,Munro_NF2016} When the reacting ions are nonrelativistic then the lowest order correction for relativistic effects is obtained by modifying the expression for $E_{3}^{*}$ in \eqref{e:4.1.1a}. Deriving a higher order relativistic version of the spectral kernel \eqref{e:4.2.3} should be feasible and should not alter the efficiency of the model. At low reaction energies ($K \lesssim 50\,keV$), the differential cross section for the DD and DT reactions is almost uniform in $\mu_{s}$. Assuming that it is uniform reduces the numerical integral in \eqref{e:4.2.3} from five- to four-dimensional, which can significantly improve the computational efficiency of the model. However, for higher reaction energies the nonuniformity of the differential cross section with $\mu_{s}$ can have a significant effect on spectral shapes, particularly if the ion distribution functions are anisotropic.

This paper has focussed on the forward problem, that is, how to calculate a neutron spectrum from a given ion distribution function. For analyzing experiments the inverse problem, how to identify the ion distribution functions given an observed neutron spectrum, is important. Theoretical investigations have demonstrated that it should be possible to identify certain properties of the distribution functions from the neutron spectra.\cite{Crilly_2021} Our model will support the further development and testing of these theories.

\begin{acknowledgments}
Sandia National Laboratories is a multimission laboratory managed and operated by National Technology \&
Engineering Solutions of Sandia, LLC, a wholly owned subsidiary of Honeywell International Inc., for the U.S.
Department of Energy's National Nuclear Security Administration under contract DE-NA0003525. This paper
describes objective technical results and analysis. Any subjective views or opinions that might be expressed in
the paper do not necessarily represent the views of the U.S. Department of Energy or the United States Government.
\end{acknowledgments}

\textbf{Data Availability}
The data that support the findings of this study are available from the corresponding author upon reasonable request.

\section{Appendix}
\subsection{The integral over $\mu_{s}$}\label{app:1}
We seek analytic solutions to the integral over $\mu_{s}$ in \eqref{e:4.1.9}. This can be done if the differential cross section is expressed using Legendre polynomials, as is often the case for differential cross section data.\cite{Drosg_1987} In particular we assume that
\begin{equation}
\frac{d\sigma}{d\Omega_{cm}} = \frac{d\sigma_{0}}{d\Omega_{cm}}\sum_{n=0}^{N}A_{n}P_{n}\left(\mu_{s}\right),\label{e:a1.1}
\end{equation}
where $P_{n}$ are the Legendre Polynomials, $A_{n}$ are the reduced Legendre coefficients, and $\frac{d\sigma_{0}}{d\Omega_{cm}}$ is the differential cross section for the reaction at the given reaction energy for $\mu_{s} = 1$. Values for $A_{i}$ and $\frac{d\sigma_{0}}{d\Omega_{cm}}$ are tabulated as a function of reaction energy $K=\frac{1}{2}m_{12} v_{r}^{2}$ in the standard formulation of reaction cross sections.\cite{Drosg_1987}

Given this formulation of the differential cross section and utilising the following representation of the Legendre polynomials
\begin{equation}
P_{n}\left(\mu_{s}\right) = 2^{n}\sum_{k=0}^{n}\left(\mu_{s}\right)^{k}\binom{n}{k}\binom{\frac{n+k-1}{2}}{n},
\end{equation}
then the integral over $\mu_{s}$ can be carried out analytically since it will be composed of a summation of integrals of the form
\begin{equation}\label{e:a1.2}
\int_{\mu_{L}}^{\mu_{U}}\frac{\left(\mu_{s}\right)^{k}}{\sqrt{-\gamma_{a}\left(\mu_{s}\right)^{2}+\gamma_{b}\mu_{s}+\gamma_{c}}}d\mu_{s},
\end{equation}
where
\begin{eqnarray}
\gamma_{a} &=& v_{cm}^{2},\label{e:a1.3a}\\
\gamma_{b} &=& 2\left(v_{1}\cos\alpha-v_{1}^{'}\right)C,\label{e:a1.3b}\\
\gamma_{c} &=& v_{1}^{2}\sin^{2}\alpha-C^{2},\label{e:a1.3c}
\end{eqnarray}

The integral \eqref{e:a1.2} can be solved by completing the square of the denominator and substitution to give
\begin{equation}\label{e:a1.4}
\frac{p^{k}}{\sqrt{\gamma_{a}}}\sum_{j=0}^{k}{k\choose j}q^{k-j}\left[F_{j}\left(x_{U}\right)-F_{j}\left(x_{L}\right)\right],
\end{equation}
where
\begin{eqnarray}
p &=& \sqrt{\frac{\gamma_{c}}{\gamma_{a}}+\left(\frac{\gamma_{b}}{2\gamma_{a}}\right)^{2}},\label{e:a1.5a}\\
q &=& \frac{\gamma_{b}}{\sqrt{4\gamma_{a}\gamma_{c}+\gamma_{b}^{2}}},\label{e:a1.5b}\\
x_{U,L} &=& \frac{1}{p}\left(\mu_{U,L}-\frac{\gamma_{b}}{2\gamma_{a}}\right).\label{e:a1.5c}
\end{eqnarray}

The function $F_{j}$ in \eqref{e:a1.4} is defined by the following indefinite integral
\begin{eqnarray}\label{e:a1.1.1}
&&F_{j}\left(x\right)=\int\frac{x^{j}}{\sqrt{1-x^{2}}}dx  =\nonumber\\
&& \begin{cases}
\sqrt{1-x^{2}}\left(a_{j-1}x^{j-1}+a_{j-3}x^{j-3}+\ldots+a_{0}\right)\quad if\quad \textrm{mod}\left(j,2\right) \neq 0,\ \\
\sqrt{1-x^{2}}\left(a_{j-1}x^{j-1}+a_{j-3}x^{j-3}+\ldots+a_{1}x\right)-a_{1}\arcsin\left(x\right)\quad if\quad \textrm{mod}\left(j,2\right)= 0,\end{cases}\nonumber
\end{eqnarray}
where the coefficients are defined recursively,
\begin{eqnarray}\label{e:a1.1.2}
a_{j-1} = \frac{-1}{j},\nonumber\\
a_{j} = \frac{j+2}{j+1}a_{j+2}.\nonumber
\end{eqnarray}
The function $F_{j}$ has the following recursive definition
\begin{equation}\label{e:a1.1.3}
F_{j}\left(x\right) =
\frac{j-1}{j}F_{j-2}\left(x\right)-\frac{1}{j}x^{j-1}\sqrt{1-x^{2}},\nonumber
\end{equation}
The first few results for this integral are given in table \ref{t:1}.
\begin{table}\centering
\caption[My table caption]{The function $F_{j}\left(x\right)$.}\label{t:1}
\begin{footnotesize}
\begin{tabular}{l l l }
\hline
\hline
 j & \quad & $F_{j}\left(x\right)$\\
 \hline
 0 & \quad & $\arcsin\left(x\right)$\\
 1 & \quad & $-\sqrt{1-x^{2}}$ \\
 2 & \quad & $-\sqrt{1-x^{2}}\left(\frac{1}{2}x\right)+\frac{1}{2}\arcsin\left(x\right)$ \\
 3 & \quad & $-\sqrt{1-x^{2}}\left(\frac{1}{3}x^{2}+\frac{2}{3}\right)$\\
 4 & \quad &  $-\sqrt{1-x^{2}}\left(\frac{1}{4}x^{3}+\frac{3}{8}x\right)+\frac{3}{8}\arcsin\left(x\right)$ \\
 5 & \quad &  $-\sqrt{1-x^{2}}\left(\frac{1}{5}x^{4}+\frac{4}{15}x^{2}+\frac{8}{15}\right)$\\
 6 & \quad &  $-\sqrt{1-x^{2}}\left(\frac{1}{6}x^{5}+\frac{5}{24}x^{3}+\frac{5}{16}x\right)+\frac{5}{16}\arcsin\left(x\right)$\\
 7 & \quad &  $-\sqrt{1-x^{2}}\left(\frac{1}{7}x^{6}+\frac{6}{35}x^{4}+\frac{8}{35}x^{2}+\frac{16}{35}\right)$\\
 8 & \quad &  $-\sqrt{1-x^{2}}\left(\frac{1}{8}x^{7}+\frac{7}{48}x^{5}+\frac{35}{192}x^{3}+\frac{35}{128}x\right)+\frac{35}{128}\arcsin\left(x\right)$\\
 9 & \quad &  $-\sqrt{1-x^{2}}\left(\frac{1}{9}x^{8}+\frac{8}{63}x^{6}+\frac{16}{105}x^{4}+\frac{64}{315}x^{2}+\frac{128}{315}\right)$\\
 10 & \quad &  $-\sqrt{1-x^{2}}\left(\frac{1}{10}x^{9}+\frac{9}{80}x^{7}+\frac{21}{160}x^{5}+\frac{21}{128}x^{3}+\frac{63}{256}x\right)+\frac{63}{256}\arcsin\left(x\right)$\\
\hline
\hline
\end{tabular}
\end{footnotesize}
\end{table}

In summary, we can express the solution to the integral over $\mu_{s}$ in \eqref{e:4.1.9} as
\begin{eqnarray}
&&\int_{\mu_{L}}^{\mu_{U}}\frac{1}{\sqrt{B^{2}-\left(C-A\right)^{2}}}\frac{d\sigma}{d\Omega_{cm}}d\mu_{s} = \nonumber\\
&&\frac{d\sigma_{0}}{d\Omega_{cm}}\sum_{n=0}^{N}\sum_{k=0}^{n}\sum_{j=0}^{k}\binom{n}{k}\binom{\frac{n+k-1}{2}}{n}\binom{k}{j} A_{n}2^{n}\frac{p^{k}}{\sqrt{\gamma_{a}}}q^{k-j}\left[F_{j}\left(x_{U}\right)-F_{j}\left(x_{L}\right)\right].
\end{eqnarray}
For DD and DT reactions, $N\sim 2$ for $K\sim 100\,keV$ and $N\sim 10$ for $K\sim10\,MeV$, which means that using the above analytic expression is significantly faster than numerical quadrature of \eqref{e:4.1.9}.

\subsection{Limiting cases for spectrum kernel}\label{app:2}
Equation \eqref{e:4.1.9} contains an expression for the spectrum kernel for arbitrary values of $v_{cm}$, $v_{r}$ and $\cos\alpha$. Here we give expressions for particular values which represent limiting cases of \eqref{e:4.1.9}.

We first consider the case of $v_{cm} \neq 0$ and $\sin\alpha = 0$.  This corresponds to vectors $\vec{v}_{1}$ and $\vec{v}_{2}$ being parallel or anti-parallel. In this case \eqref{e:4.1.3c} is reduced to
\begin{equation}\label{e:a2.1}
\mu' = \pm \mu_{s}.
\end{equation}

This means $E_{3}$ is independent of $\phi_{s}$, and so we can express $\mu_{s}$ as
\begin{equation}\label{e:a2.2}
\mu_{s}= \frac{E_{3}^{*}-E_{3}+\frac{1}{2}m_{3}v_{cm}^{2}}{v_{cm}\sqrt{2m_{3}E_{3}^{*}}},
\end{equation}
which results in the following expression for the spectrum kernel
\begin{equation}\label{e:a2.3}
\mathcal{K}_{12}\left(E_{3};v_{cm},v_{r}\right) = 2\pi\frac{1}{v_{cm}}\frac{1}{\sqrt{2m_{3}E_{3}^{*}}}\frac{d\sigma}{d\Omega_{cm}}dE_{3}.
\end{equation}
Note how integration over $\mu_{s}$ is not required in this case since we can use \eqref{e:a2.2} to calculate the correct value of $\frac{d\sigma}{d\Omega_{cm}}$.

The second limiting case is when $v_{cm} = 0$. This corresponds to $m_{1}\vec{v}_{1} = -m_{2}\vec{v}_{2}$, resulting in all neutrons produced with an energy of $E_{3} = E_{3}^{*}$ in the lab frame. The transformation $E_{3}^{'}\rightarrow E_{3}$ and integration over $d\Omega_{cm}$ are trivial. The spectral kernel becomes
\begin{equation}\label{e:a2.4}
 \mathcal{K}_{12}\left(E_{3};v_{r}\right) = \sigma\delta\left(E_{3}-E_{3}^{*}\right)dE_{3},
\end{equation}
where $\sigma$ is the total reaction cross section ($\sigma = \int \frac{d\sigma}{d\Omega}d\Omega_{cm}$).

Equations \eqref{e:a2.3} or \eqref{e:a2.4} can replace \eqref{e:4.1.9} as an expression for the spectrum kernel in the scenario in which $v_{cm} \neq 0$, $\sin\alpha = 0$ or $v_{cm} = 0$, respectively.

\nocite{*}
\bibliographystyle{unsrt}


\begin{thebibliography}{10}

\bibitem{Munro_NF2016}
David~H. Munro.
\newblock Interpreting inertial fusion neutron spectra.
\newblock {\em Nuclear Fusion}, 56(3):036001, feb 2016.

\bibitem{Higginson_JCP2019}
Drew~Pitney Higginson, Anthony Link, and Andrea Schmidt.
\newblock A pairwise nuclear fusion algorithm for weighted particle-in-cell
  plasma simulations.
\newblock {\em Journal of Computational Physics}, 388:439 -- 453, 2019.

\bibitem{Taitano_POP2018}
W.~T. Taitano, A.~N. Simakov, L.~Chacon, and B.~Keenan.
\newblock Yield degradation in inertial-confinement-fusion implosions due to
  shock-driven kinetic fuel-species stratification and viscous heating.
\newblock {\em Physics of Plasmas}, 25(5):056310, 2018.

\bibitem{Crilly_POP2018}
A.~J. Crilly, B.~D. Appelbe, K.~McGlinchey, C.~A. Walsh, J.~K. Tong, A.~B.
  Boxall, and J.~P. Chittenden.
\newblock Synthetic nuclear diagnostics for inferring plasma properties of
  inertial confinement fusion implosions.
\newblock {\em Physics of Plasmas}, 25(12):122703, 2018.

\bibitem{Crilly_POP2020}
A.~J. Crilly, B.~D. Appelbe, O.~M. Mannion, C.~J. Forrest, V.~Gopalaswamy,
  C.~A. Walsh, and J.~P. Chittenden.
\newblock Neutron backscatter edge: A measure of the hydrodynamic properties of
  the dense dt fuel at stagnation in icf experiments.
\newblock {\em Physics of Plasmas}, 27(1):012701, 2020.

\bibitem{Crilly_POP2021}
A.~J. Crilly, B.~D. Appelbe, O.~M. Mannion, C.~J. Forrest, and J.~P.
  Chittenden.
\newblock The effect of areal density asymmetries on scattered neutron spectra
  in icf implosions.
\newblock {\em Physics of Plasmas}, 28(2):022710, 2021.

\bibitem{Frenje_PPCF2020}
J~A Frenje.
\newblock Nuclear diagnostics for inertial confinement fusion ({ICF}) plasmas.
\newblock {\em Plasma Physics and Controlled Fusion}, 62(2):023001, jan 2020.

\bibitem{Frenje_NF2013}
J.A. Frenje, R.~Bionta, E.J. Bond, J.A. Caggiano, D.T. Casey, C.~Cerjan,
  J.~Edwards, M.~Eckart, D.N. Fittinghoff, S.~Friedrich, V.Yu. Glebov,
  S.~Glenzer, G.~Grim, S.~Haan, R.~Hatarik, S.~Hatchett, M.~Gatu Johnson, O.S.
  Jones, J.D. Kilkenny, J.P. Knauer, O.~Landen, R.~Leeper, S.~Le Pape,
  R.~Lerche, C.K. Li, A.~Mackinnon, J.~McNaney, F.E. Merrill, M.~Moran, D.H.
  Munro, T.J. Murphy, R.D. Petrasso, R.~Rygg, T.C. Sangster, F.H. S{\'{e}}guin,
  S.~Sepke, B.~Spears, P.~Springer, C.~Stoeckl, and D.C. Wilson.
\newblock Diagnosing implosion performance at the national ignition facility
  ({NIF}) by means of neutron spectrometry.
\newblock {\em Nuclear Fusion}, 53(4):043014, mar 2013.

\bibitem{GatuJohnson_RSI2012}
M.~Gatu Johnson, J.~A. Frenje, D.~T. Casey, C.~K. Li, F.~H. Séguin,
  R.~Petrasso, R.~Ashabranner, R.~M. Bionta, D.~L. Bleuel, E.~J. Bond, J.~A.
  Caggiano, A.~Carpenter, C.~J. Cerjan, T.~J. Clancy, T.~Doeppner, M.~J.
  Eckart, M.~J. Edwards, S.~Friedrich, S.~H. Glenzer, S.~W. Haan, E.~P.
  Hartouni, R.~Hatarik, S.~P. Hatchett, O.~S. Jones, G.~Kyrala, S.~Le~Pape,
  R.~A. Lerche, O.~L. Landen, T.~Ma, A.~J. MacKinnon, M.~A. McKernan, M.~J.
  Moran, E.~Moses, D.~H. Munro, J.~McNaney, H.~S. Park, J.~Ralph, B.~Remington,
  J.~R. Rygg, S.~M. Sepke, V.~Smalyuk, B.~Spears, P.~T. Springer, C.~B.
  Yeamans, M.~Farrell, D.~Jasion, J.~D. Kilkenny, A.~Nikroo, R.~Paguio, J.~P.
  Knauer, V.~Yu~Glebov, T.~C. Sangster, R.~Betti, C.~Stoeckl, J.~Magoon, M.~J.
  Shoup, G.~P. Grim, J.~Kline, G.~L. Morgan, T.~J. Murphy, R.~J. Leeper, C.~L.
  Ruiz, G.~W. Cooper, and A.~J. Nelson.
\newblock Neutron spectrometry—an essential tool for diagnosing implosions at
  the national ignition facility (invited).
\newblock {\em Review of Scientific Instruments}, 83(10):10D308, 2012.

\bibitem{Mannion_POP2021}
O.~M. Mannion, I.~V. Igumenshchev, K.~S. Anderson, R.~Betti, E.~M. Campbell,
  D.~Cao, C.~J. Forrest, M.~Gatu Johnson, V.~Yu. Glebov, V.~N. Goncharov,
  V.~Gopalaswamy, S.~T. Ivancic, D.~W. Jacobs-Perkins, A.~Kalb, J.~P. Knauer,
  J.~Kwiatkowski, A.~Lees, F.~J. Marshall, M.~Michalko, Z.~L. Mohamed,
  D.~Patel, H.~G. Rinderknecht, R.~C. Shah, C.~Stoeckl, W.~Theobald, K.~M. Woo,
  and S.~P. Regan.
\newblock Mitigation of mode-one asymmetry in laser-direct-drive inertial
  confinement fusion implosions.
\newblock {\em Physics of Plasmas}, 28(4):042701, 2021.

\bibitem{GatuJohnson_PRE2016}
M.~Gatu~Johnson, J.~P. Knauer, C.~J. Cerjan, M.~J. Eckart, G.~P. Grim, E.~P.
  Hartouni, R.~Hatarik, J.~D. Kilkenny, D.~H. Munro, D.~B. Sayre, B.~K. Spears,
  R.~M. Bionta, E.~J. Bond, J.~A. Caggiano, D.~Callahan, D.~T. Casey,
  T.~D\"oppner, J.~A. Frenje, V.~Yu. Glebov, O.~Hurricane, A.~Kritcher,
  S.~LePape, T.~Ma, A.~Mackinnon, N.~Meezan, P.~Patel, R.~D. Petrasso, J.~E.
  Ralph, P.~T. Springer, and C.~B. Yeamans.
\newblock Indications of flow near maximum compression in layered
  deuterium-tritium implosions at the national ignition facility.
\newblock {\em Phys. Rev. E}, 94:021202, Aug 2016.

\bibitem{Murphy_2014}
T.~J. Murphy.
\newblock The effect of turbulent kinetic energy on inferred ion temperature
  from neutron spectra.
\newblock {\em Physics of Plasmas}, 21(7):072701, 2014.

\bibitem{Brysk_1973}
H~Brysk.
\newblock Fusion neutron energies and spectra.
\newblock {\em Plasma Physics}, 15(7):611--617, jul 1973.

\bibitem{Ballabio_1998}
L~Ballabio, J~Källne, and G~Gorini.
\newblock Relativistic calculation of fusion product spectra for thermonuclear
  plasmas.
\newblock {\em Nuclear Fusion}, 38(11):1723--1735, nov 1998.

\bibitem{Appelbe_2011}
B~Appelbe and J~Chittenden.
\newblock The production spectrum in fusion plasmas.
\newblock {\em Plasma Physics and Controlled Fusion}, 53(4):045002, feb 2011.

\bibitem{Appelbe_HEDP2014}
B.~Appelbe and J.~Chittenden.
\newblock Relativistically correct dd and dt neutron spectra.
\newblock {\em High Energy Density Physics}, 11:30--35, 2014.

\bibitem{Munro_POP2017}
D.~H. Munro, J.~E. Field, R.~Hatarik, J.~L. Peterson, E.~P. Hartouni, B.~K.
  Spears, and J.~D. Kilkenny.
\newblock Impact of temperature-velocity distribution on fusion neutron peak
  shape.
\newblock {\em Physics of Plasmas}, 24(5):056301, 2017.

\bibitem{Appelbe_POP2012}
B.~Appelbe and J.~Chittenden.
\newblock Quasi-monoenergetic spectra from reactions in a beam-target plasma.
\newblock {\em Physics of Plasmas}, 19(7):073115, 2012.

\bibitem{Appelbe_POP2015}
B.~Appelbe and J.~Chittenden.
\newblock Neutron spectra from beam-target reactions in dense z-pinches.
\newblock {\em Physics of Plasmas}, 22(10):102703, 2015.

\bibitem{Knapp_POP2013}
P.~F. Knapp, D.~B. Sinars, and K.~D. Hahn.
\newblock Diagnosing suprathermal ion populations in z-pinch plasmas using
  fusion neutron spectra.
\newblock {\em Physics of Plasmas}, 20(6):062701, 2013.

\bibitem{Higginson_POP2019}
D.~P. Higginson, J.~S. Ross, D.~D. Ryutov, F.~Fiuza, S.~C. Wilks, E.~P.
  Hartouni, R.~Hatarik, C.~M. Huntington, J.~Kilkenny, B.~Lahmann, C.~K. Li,
  A.~Link, R.~D. Petrasso, B.~B. Pollock, B.~A. Remington, H.~G. Rinderknecht,
  Y.~Sakawa, H.~Sio, G.~F. Swadling, S.~Weber, A.~B. Zylstra, and H.-S. Park.
\newblock Kinetic effects on neutron generation in moderately collisional
  interpenetrating plasma flows.
\newblock {\em Physics of Plasmas}, 26(1):012113, 2019.

\bibitem{Klir_NJP2020}
D~Klir, AV~Shishlov, VA~Kokshenev, SL~Jackson, K~Rezac, RK~Cherdizov,
  J~Cikhardt, GN~Dudkin, FI~Fursov, J~Krasa, et~al.
\newblock Production of energetic protons, deuterons, and neutrons up to 60 mev
  via disruption of a current-carrying plasma column at 3 ma.
\newblock {\em New Journal of Physics}, 22(10):103036, 2020.

\bibitem{Mannion_2021}
O. M. Mannion, W. T. Taitano, B. D. Appelbe, A. J. Crilly, C. J. Forrest, V. Yu.
Glebov, J. P. Knauer, P. W. McKenty, Z. L. Mohamed, C. Stoeckl, B. D. Keenan,
J. P. Chittenden, P. Adrian, J. Frenje, N. Kabadi, M. Gatu Johnson, and S. P. Regan
\newblock Evidence of Non-Maxwellian Ion Velocity Distributions in Spherical Shock Driven
    Implosions.
\newblock {\em Physical Review E}, (submitted).

\bibitem{Inglebert_EPL2014}
A.~Inglebert, B.~Canaud, and O.~Larroche.
\newblock Species separation and modification of neutron diagnostics in
  inertial-confinement fusion.
\newblock {\em {EPL} (Europhysics Letters)}, 107(6):65003, sep 2014.

\bibitem{Anderson_JCP2020}
S.E. Anderson, W.T. Taitano, L.~Chacon, and A.N. Simakov.
\newblock An efficient, conservative, time-implicit solver for the fully
  kinetic arbitrary-species 1d-2v vlasov-ampère system.
\newblock {\em Journal of Computational Physics}, 419:109686, 2020.

\bibitem{Taitano_CPC2021}
W.T. Taitano, B.D. Keenan, L.~Chacon, S.E. Anderson, H.R. Hammer, and A.N.
  Simakov.
\newblock An eulerian vlasov-fokker–planck algorithm for spherical implosion
  simulations of inertial confinement fusion capsules.
\newblock {\em Computer Physics Communications}, 263:107861, 2021.

\bibitem{Keenan_POP2020}
Brett~D. Keenan, William~T. Taitano, and Kim Molvig.
\newblock Physics of the implosion up until the time of ignition in a revolver
  (triple-shell) capsule.
\newblock {\em Physics of Plasmas}, 27(4):042704, 2020.

\bibitem{Keenan_POP2018}
Brett~D. Keenan, Andrei~N. Simakov, William~T. Taitano, and Luis Chacón.
\newblock Ion species stratification within strong shocks in two-ion plasmas.
\newblock {\em Physics of Plasmas}, 25(3):032103, 2018.

\bibitem{Chacon_JCP2017}
L.~Chacon, G.~Chen, D.A. Knoll, C.~Newman, H.~Park, W.~Taitano, J.A. Willert,
  and G.~Womeldorff.
\newblock Multiscale high-order/low-order (holo) algorithms and applications.
\newblock {\em Journal of Computational Physics}, 330:21--45, 2017.

\bibitem{Goncharov_2015}
P.R. Goncharov.
\newblock Spectra of neutrons from a beam-driven fusion source.
\newblock {\em Nuclear Fusion}, 55(6):063012, may 2015.

\bibitem{Drosg_1987}
M~Drosg and O~Schwerer.
\newblock Production of monoenergetic neutrons between 0.1 and 23 mev. neutron
  energies and cross-sections.
\newblock In {\em Handbook on nuclear activation data}. 1987.

\bibitem{Appelbe_PPCF2011}
B~Appelbe and J~Chittenden.
\newblock The production spectrum in fusion plasmas.
\newblock {\em Plasma Physics and Controlled Fusion}, 53(4):045002, feb 2011.

\bibitem{Crilly_2021}
A.~J. Crilly, B.~D. Appelbe, O.~M. Mannion, W.~T. Taitano, E.~P. Hartouni,
  A.~S. Moore, M.~Gatu-Johnson, and J.~P. Chittenden.
\newblock Constraints on ion velocity distributions from fusion product
  spectroscopy.
\newblock {\em Nuclear Fusion}, 62:126015, 2022.

\bibitem{Hartouni_2023}
E. P. Hartouni, A. S. Moore, A. J. Crilly, B. D. Appelbe, P. A. Amendt, K. L. Baker, D. T. Casey, D. S. Clark, T. D{\"o}ppner, M. J. Eckart,  and et~al.
\newblock Evidence for suprathermal ion distribution in burning plasmas
\newblock {\em Nature Physics}, 19(1):72--77, 2023.

\end{thebibliography}



\end{document}